\documentclass[pra, twocolumn, showkeys, floatfix]{revtex4}
\usepackage{graphicx}
\usepackage{color}
\usepackage{amsmath, amsfonts, amssymb, bm}
\usepackage{tikz}
\usepackage{slashed}
\newcommand{\dd}{\text{d}}
\usepackage[]{psfrag}
\usetikzlibrary{decorations.pathmorphing}

\begin{document}
	\title{Comparative study of nonperturbative electron-positron pair production by intense laser 
	fields colliding with either bremsstrahlung $\gamma$-rays or relativistic ions}

	\author{S.~Remme}
	\author{A.~B.~Voitkiv}
	\author{S.~Villalba-Ch\'avez}
	\author{C.~M\"uller}

	\affiliation{Institut f\"ur Theoretische Physik I, Heinrich-Heine-Universit\"at D\"usseldorf, Universit\"atsstra{\ss}e~1, 40225 D\"usseldorf, Germany}
	\date{\today}
	\begin{abstract}
It is well known that electron-positron pairs can be created in the strong-field environments formed by (i) a high-intensity laser field and a high-energy $\gamma$-photon (nonlinear Breit-Wheeler process) or (ii) a high-intensity laser field and a nuclear Coulomb field (nonlinear Bethe-Heitler process). Both of these processes are particularly interesting in the interaction regime where the laser field enters nonperturbatively. Various experimental collaborations are currently aiming at detecting for the first time the nonperturbative Breit-Wheeler process, by exploiting high-intensity laser pulses and $\gamma$-ray sources based on bremsstrahlung. In contrast, an experimental observation of the nonlinear, nonperturbative Bethe-Heitler process still lies further ahead in the future because its technical realization appears at present more challenging. Our comparative study shows, however, that the physical properties of the total rates for the processes (i) and (ii) can become remarkably similar when the parameters for the bremsstrahlung-driven nonperturbative Breit-Wheeler process are properly chosen. In this sense, the upcoming experiments on the nonlinear Breit-Wheeler process could also be used to closely ``simulate'' the currently hard to observe nonlinear Bethe-Heitler process.
	\end{abstract}

	\maketitle
	
	\section{Introduction}
Soon after the construction of the first laser in 1960, theoreticians started to consider quantum electrodynamic processes that could arise in laser fields of very high intensity. In particular, various possibilities of field-induced electron-positron pair production were investigated. Since a plane laser wave alone cannot extract pairs from the quantum vacuum \cite{Schwinger}, it needs to be combined with an additional field. A suitable candidate for the latter is the field of a high-energy $\gamma$-photon $\omega'$, which enables pair production via the nonlinear Breit-Wheeler process \cite{Reiss-1962, Nikishov-Ritus-1964, Ritus-Review} according to the reaction
\begin{eqnarray}
\label{BW}
n\omega + \omega' \to e^+e^-\,.
\end{eqnarray}
Here, $\omega$ denotes the laser frequency and $n$ the number of absorbed laser photons. 
Another suitable candidate is the Coulomb field of a bare ion of charge number $Z$. When combined with a strong laser field, the nonlinear Bethe-Heitler process \cite{Yakovlev} can occur
\begin{eqnarray}
\label{BH}
n\omega + Z \to Z + e^+e^-\,.
\end{eqnarray}
If the ion moves relativistically at high Lorentz factor $\gamma$, the laser frequency and laser field strength are enhanced in the ionic rest frame by the Lorentz boost. Various interactions regimes of the processes \eqref{BW} and \eqref{BH} exist, which are basically distinguished by the dimensionless intensity parameter $\xi = eF/(mc\omega)$, where $F$ denotes the electric field amplitude of the laser.

Motivated by the ongoing progress in high-intensity laser technology, the nonlinear processes \eqref{BW} and \eqref{BH} have been studied intensively by theoreticians during the last two decades (see \cite{Review1, Review2, Review3, Review4} for reviews). A wide variety of physical scenarios was considered, including pulsed \cite{Heinzl2010, Lebed, Krajewska-BW, Titov2012, Krajewska-PRA2013, Meuren-DiPiazza, Jansen, DiPiazzaFocus, Heinzl2020, King2021, Krajewska-NJP} and bichromatic laser fields \cite{Narozhny2, Loetstedt, DiPiazza-PRL, Krajewska-PRA2012, Augustin, Roshchupkin, Mahlin2023}, different polarization configurations \cite{Avetissian, MVG-PRA2003, Sieczka, Krajewska-PRA2006, Milstein, Kuchiev, DiPiazza-PLB, Titov2020, Hatsagortsyan-PRR, Riconda, Seipt2020, Hatsagortsyan-PRD, Podszus, Seipt}, spin effects \cite{Seipt2020, Hatsagortsyan-PRD, Podszus, Seipt, Serbo, DiPiazza-Spin, Tim-Oliver, Selym, Jansen-Spin, Tang-Spin} and laser-driven recollisions of the created electron-positron pair \cite{Recol1, Recol2}. For the nonlinear Bethe-Heitler process, also the relevance of bound atomic states and the impact of multiple Coulomb centers were analyzed \cite{MVG-PRL, Deneke, diatomic, Grobe-bound, Grobe-phase, Remme}. 

Concerning experimental studies, the nonlinear Breit-Wheeler effect has been observed in a few-photon regime at a moderate value of $\xi\lesssim 1$, where the production rate was found to follow a power law of the form $R\sim \xi^{2n}$ \cite{SLAC}. The experiment relied on Compton scattering in highly relativistic electron-laser collisions, followed by the pair-producing reaction \eqref{BW}. At present, the complementary and hitherto unobserved regime of pair production with $\xi\gg 1$ is of special interest where the laser field couples to the particles in a manifestly nonperturbative way. The first observation of the nonperturbative Breit-Wheeler process is currently coming into experimental reach; corresponding campaigns have been proposed at several laboratories \cite{ELI, CoReLS, FACET, Gemini, LUXE, CALA}. A promising route to create the required high-energy $\gamma$-rays relies on bremsstrahlung from a relativistic incident electron beam \cite{LUXE, CALA, Reiss1971, Blackburn2018, Hartin, Eckey2022, Golub2022, King2024, Eckey2024, Elsner, MacLeod}. In contrast, the prospects for an experimental detection of the nonlinear Bethe-Heitler process in the near future are less bright since it would require to merge a powerful ion accelarator with a high-intensity laser source, a combination which at present does not exist.

In this paper, our goal is to show that the upcoming discovery experiments of the nonperturbative Breit-Wheeler process \eqref{BW} could, in principle, also be used to 'mimick' the nonperturbative Bethe-Heitler effect \eqref{BH}. Both processes are formaly related by the fact that pairs are created in a laser field by absorption of a {\it real} photon $\omega'$ in the former, and a {\it virtual} photon from the nuclear Coulomb field in the latter case. As we will demonstrate, this connection becomes particularly close when the nonperturbative Breit-Wheeler process involves $\gamma$-photons from a bremsstrahlung source, instead of a monoenergetic $\gamma$-beam, because the associated energy spectrum of the bremsstrahlung $\gamma$-photons is very broad and falls off essentially with $1/\omega'$ \cite{Jackson}. Similarly, in the spirit of the Weizs\"acker-Williams method \cite{Landau}, the electromagnetic field of a relativistically moving ion may be decomposed into a broad spectrum of equivalent photons whose frequency dependence resembles the bremsstrahlung spectrum. As a result, the similarity between the nonlinear Breit-Wheeler and Bethe-Heitler processes can become very intimate, leading to nearly identical pair production rates when the bremsstrahlung and nuclear beam parameters are suitably chosen.

To provide a basis for our comparative study, we will obtain analytical rate expressions for bremsstrahlung-driven nonlinear Breit-Wheeler pair production in various field scenarios. Moreover, we derive an improved analytical formula for the nonlinear Bethe-Heitler rate in a linearly polarized laser wave of very high intensity. The rate formulas will allow us to show that the rates for the nonlinear Breit-Wheeler and the nonlinear Bethe-Heitler process in the same laser field can be brought to coincide by suitably adjusting the thickness of the bremsstrahlung target and properly chosing the Lorentz factors of the incident nuclear beam and bremsstrahlung-generating electron beam, respectively.

Our paper is organized as follows. Sec.~\ref{sec: comparison} presents a comparison of the Bethe-Heitler and Breit-Wheeler rates, with the Breit-Wheeler rate being averaged over the bremsstrahlung spectrum. Sec.~\ref{subsec: undercritical} starts with the comparison of these rates in the undercritical regime, where $F' \ll F_{\rm cr}$. Here, $F'$ denotes the effective laser field amplitude which is relevant for the process and  $F_{\rm cr} = m^2 c^3 / (e \hbar)$ is the critical field of QED. Both circularly and linearly polarized laser fields are considered. Afterwards, Sec.~\ref{subsec: overcritical} discusses the comparison of both rates in the overcritical regime with $F' \gg F_{\rm cr}$. The results regarding the comparison of the pair production rates as well as the associated spectra of the equivalent photons and the bremsstrahlung are considered in Sec.~\ref{sec: results}. A conclusion is given in Sec.~\ref{sec: conclusion}.

Relativistic units with the reduced Planck constant $\hbar=1$, $4 \pi \varepsilon_0 = 1$ and the speed of light $c=1$ are used throughout. 

\section{Comparison of Breit-Wheeler and Bethe-Heitler pair production rates}
\label{sec: comparison}

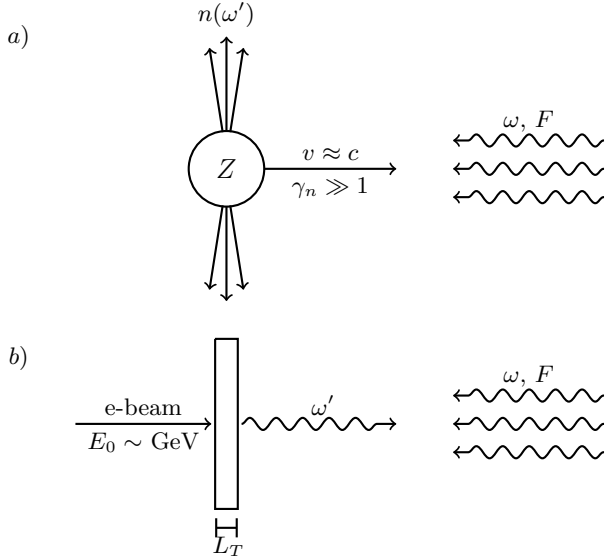
\begin{figure}[t]
	\begin{tikzpicture}[scale =0.5]
		\path[black, thick] (-5.5,3.5) node {$a)$};
		\draw[black, thick] (0,0) circle [radius=1] node[pos=-0.05] {\huge $Z$} ;
		\draw[black, thick, ->] (0,1) -- (0,3.5)node[above, pos=1] {$n(\omega')$} ;
		\draw[black, thick, ->] (-0.1,0.97) -- (-0.45,3.2);
		\draw[black, thick, ->] (0.1,0.97) -- (0.45,3.2);
		\draw[black, thick, ->] (0,-1) -- (0,-3.5);
		\draw[black, thick, ->] (-0.1,-0.97) -- (-0.45,-3.2);
		\draw[black, thick, ->] (0.1,-0.97) -- (0.45,-3.2);
		\draw[black, thick, ->] (1,0) -- (4.5,0) node[above, pos=0.5] {$v \approx c$} node[below, pos=0.5] {$\gamma_n \gg 1$};
		\draw [black, thick,<-,decorate,decoration={snake,amplitude=.8mm,segment length=3.5mm,pre length=2mm}] (6,0.75) -- (10,0.75) node[above, pos=0.5] {$\omega$, $F$};
		\draw [black, thick,<-,decorate,decoration={snake,amplitude=.8mm,segment length=3.5mm,pre length=2mm}] (6,0) -- (10,0);
		\draw [black, thick,<-,decorate,decoration={snake,amplitude=.8mm,segment length=3.5mm,pre length=2mm}] (6,-0.75) -- (10,-0.75);
		\path[black, thick] (-5.5,-5) node {$b)$};
		\draw[black, thick] (-0.3,-4.5) -- (0.3,-4.5)--(0.3,-9)--(-0.3,-9)--(-0.3,-4.5);
		\draw[black, thick, |-|] (0.3,-9.5)--(-0.3,-9.5) node[below,pos=0.4] {$L_T$};
		\draw[black, thick, ->] (-4,-6.75) -- (-0.4,-6.75) node[above, pos=0.5] {e-beam} node[below, pos=0.5] {$E_0 \sim$ GeV};
		\draw [black, thick,->,decorate,decoration={snake,amplitude=.8mm,segment length=3.5mm,post length=2mm}] (0.4,-6.75) -- (4.5,-6.75) node[above, pos=0.5] {\ $\omega'$};
		\draw [black, thick,<-,decorate,decoration={snake,amplitude=.8mm,segment length=3.5mm,pre length=2mm}] (6,-6) -- (10,-6) node[above, pos=0.5] { $\omega$, $F$};
		\draw [black, thick,<-,decorate,decoration={snake,amplitude=.8mm,segment length=3.5mm,pre length=2mm}] (6,-6.75) -- (10,-6.75);
		\draw [black, thick,<-,decorate,decoration={snake,amplitude=.8mm,segment length=3.5mm,pre length=2mm}] (6,-7.5) -- (10,-7.5);
	\end{tikzpicture}
	\caption[justification=justified]{Schematic illustration of pair production processes. Panel a) illustrates the nonlinear Bethe-Heitler process where a highly relativistic nucleus with charge $Z$ and equivalent photon spectrum $n (\omega ')$ interacts with a strong laser field of frequency $\omega$ and field strength $F$. In panel b) the nonlinear Breit-Wheeler process is shown. Here, an electron beam with energies $E_0$ of a few GeV collides with a target of thickness $L_T$ such that bremsstrahlung $\omega '$ is produced. The electron-positron pair is created through the collision of the bremsstrahlung $\omega'$ with a strong laser field.}
	\label{fig:schema}
\end{figure}

In the nonlinear Bethe-Heitler pair production process, the electron-positron pair is created in the presence of a bare ion (for illustration see panel a) of Fig.~\ref{fig:schema}). If the ion moves at relativistic velocity, its electromagnetic field can be decomposed into a spectrum of equivalent photons through \cite{Jackson}
\begin{equation}
	\label{eq: spectrum equivalent photon_exact}
	n (\omega') = \frac{2}{\pi} \frac{Z^2 \alpha}{\omega'} \left[x \text{K}_0(x) \text{K}_1(x) - \frac{x^2}{2} \left(\text{K}_1^2(x) - \text{K}_0^2(x) \right) \right].
\end{equation}
Here, $Z$ denotes the nuclear charge of the ion, $\alpha$ the fine-structure constant, $\omega'$ the equivalent photon frequency, $\text{K}_0$ and $\text{K}_1$ are the modified Bessel-functions of second kind and $x= \frac{\omega'}{\gamma_n m}$ with the Lorentz-factor $\gamma_n$ and the electron mass $m$. For small values of $x$, hence for large nuclear Lorentz-factors $\gamma_n$, the equivalent photon spectrum can be approximated by \cite{Landau}
\begin{equation}
	\label{eq: spectrum equivalent photon}
	n (\omega') = \frac{2}{\pi} \frac{Z^2 \alpha}{\omega'} \ln \left(\frac{\gamma_n m}{\omega'}\right) \sim \frac{1}{\omega'}.
\end{equation}

In comparison, we consider the nonlinear Breit-Wheeler pair production process, where the electron-positron pair is created through a high-energy $\gamma$-photon and a laser field, as illustrated in the second panel of Fig.~\ref{fig:schema}. We consider the high-energy $\gamma$-photon to be resulting from a bremsstrahlung source. Its energy spectrum for thin targets reads \cite{bremsstrahlung1, bremsstrahlung2}
\begin{equation}
	\label{eq: spectrum bremsstrahlung}
	I_{\gamma} (\omega',\ell) = \frac{\ell}{\omega'} \left(\frac{4}{3} - \frac{4 \omega'}{3 E_0} + \left(\frac{\omega'}{E_0}\right)^2 \right) \sim \frac{1}{\omega'},
\end{equation}
with the bremsstrahlung energy $\omega'$, the energy $E_0$ of the incident electrons (corresponding to a Lorentz factor $\gamma_e = \frac{E_0}{m}$) and the normalized target thickness $\ell = \frac{L_{\rm T}}{L_{\rm rad}}$, where $L_{\rm rad}$ is the target-specific radiation length.

The dependence of the bremsstrahlung spectrum on the frequency $\omega'$ resembles the dependence of the equivalent photon spectrum. This similarity suggests that the nonlinear Breit-Wheeler process involving bremsstrahlung photons could 'mimic' the nonlinear Bethe-Heitler process. In the following Section, we therefore compare the resulting pair production rates for both processes, where the rate of the Breit-Wheeler pair production is averaged over the energy spectrum \eqref{eq: spectrum bremsstrahlung} of the bremsstrahlung $\gamma$-photons.

\subsection{Nonperturbative pair production in the undercritical regime}
\label{subsec: undercritical}

\subsubsection{Circularly polarized laser field}
\label{subsubsec: undercritical_circ}
For field strengths $F'$ smaller than the critical field $F_{\rm cr}$, the pair production process occurs through a tunneling procedure, resulting in an exponential dependence on the applied laser field strength $F$. For a circularly polarized laser field the Bethe-Heitler pair production rate in the laboratory frame is given by \cite{Milstein}
\begin{equation}
	\label{eq: RBHcirc_under}
	\mathcal{R}_{\rm BH,circ}^{\rm (uc)} = \frac{1}{\gamma_n} \frac{(Z \alpha)^2}{2 \sqrt{\pi}} m \left(\frac{\eta}{2 \sqrt{3}}\right)^{5/2} \exp \left(- \frac{2 \sqrt{3}}{\eta}\right).
\end{equation}
Here, $\eta = \frac{F'}{F_{\rm cr}}$ with the laser amplitude $F' = (1 + \beta_n) \gamma_n \, F$ in the nuclear rest frame, $F$ being the laser field amplitude in the laboratory frame and $\beta_n = \frac{v}{c}$.

We want to compare the Bethe-Heitler pair production rate \eqref{eq: RBHcirc_under} with the rate of the Breit-Wheeler process in a circularly polarized field averaged over the bremsstrahlung spectrum \eqref{eq: spectrum bremsstrahlung}. For a given frequency $\omega '$, the Breit-Wheeler rate reads \cite{Ritus-Review}
\begin{equation}
	\label{eq: RBWcirc_under}
	\mathcal{R}_{\rm BW,circ}^{\rm (uc)} = \frac{3}{2^4} \sqrt{\frac{3}{2}} \frac{\alpha m^2}{\omega'} \kappa \exp \left( - \frac{8}{3 \kappa} \right),
\end{equation}
with $\kappa=\frac{F'}{F_{\rm cr}}$, where $F' = \frac{2 \omega'}{m} F$ denotes the boosted laser field amplitude assuming that the latter and $\gamma$ beams collide head-on. We point out that the numerical factor $8/3$ appearing in the exponential of Eq.~\eqref{eq: RBWcirc_under} is smaller than the corresponding factor $2 \sqrt{3}$ in Eq.~\eqref{eq: RBHcirc_under}.

Averaging the Breit-Wheeler rate \eqref{eq: RBWcirc_under} over the bremsstrahlung spectrum \eqref{eq: spectrum bremsstrahlung} yields
\begin{equation}
\label{eq: RBWcirc_under_av1}
\overline{\mathcal{R}}_{\rm BW,circ}^{\rm (uc)} = \int_0^{E_0} I_{\gamma} (\omega',\ell)  \mathcal{R}_{\rm BW,circ}^{\rm (uc)} (\omega') \, \dd \omega'.
\end{equation}
It describes the rate associated with an incident electron that emits radiation.\\
Analytic calculation of the integral in \eqref{eq: RBWcirc_under_av1} results in
\begin{equation}
\label{eq: RBWcirc_under_av2}
\begin{split}
& \overline{\mathcal{R}}_{\rm BW,circ}^{\rm (uc)} = \frac{3}{2^4} \sqrt{\frac{3}{2}} \alpha m^2 \frac{\ell}{E_0} \Biggl( - \text{Ei} \left(-\frac{8}{3 \kappa_0} \right) \\
& \ \ \ \times \left[\frac{\kappa_0}{2} + \frac{4}{3}+\frac{4}{3 \kappa_0}\right]- \text{e }^{-\frac{8}{3 \kappa_0}}\left[\frac{5 \kappa_0}{16} + \frac{1}{2} \right] \Biggr)
\end{split}
\end{equation}
with $\kappa_0= \frac{2E_0}{m} \frac{F}{F_{\rm cr}}$ and the exponential integral function $\text{Ei} (x) = \int_{-\infty}^{x} \frac{e^t}{t} \dd t$. As $F' \ll F_{\rm cr}$ holds in the undercritical regime, Eq.~\eqref{eq: RBWcirc_under_av2} can be expanded for small values of $\kappa_0$, leading to a compact form for $\overline{\mathcal{R}}_{\rm BW,circ}^{\rm (uc)}$ given by
\begin{equation}
\label{eq: RBWcirc_under_av}
\overline{\mathcal{R}}_{\rm BW,circ}^{\rm (uc)} \approx \frac{3^2}{2^5} \sqrt{\frac{3}{2}} \alpha \, E_0 \, \ell \left( \frac{F}{F_{\rm cr}} \right)^2 \exp \left( - \frac{8}{3 \kappa_0} \right).
\end{equation}
Note, that this averaged Breit-Wheeler rate was also obtained in \cite{Hartin}. Eq.~\eqref{eq: RBWcirc_under_av} contains the same numerical factor $8/3$ in the exponential as in Eq.~\eqref{eq: RBWcirc_under}, when in the latter the bremsstrahlung endpoint energy $\omega' = E_0$ is inserted. In the tunnel regime, the main dependence of both the Bethe-Heitler and the averaged Breit-Wheeler pair production rates on the laser field strength $F$ lies in the exponential function. In order to obtain rates of similar magnitude, we require the exponents of \eqref{eq: RBHcirc_under} and \eqref{eq: RBWcirc_under_av} to be equal. Through this, a suitable Lorentz-factor $\gamma_n$ for the nucleus in the Bethe-Heitler process can be found as
\begin{equation}
\label{eq: gamma_under}
\gamma_n \approx \frac{3 \sqrt{3}}{4} \frac{E_0}{m} \approx 1.3 \, \frac{E_0}{m}
\end{equation}
which results from the reduced velocity
\begin{equation}
	\label{eq: beta_under}
	\beta_n = \frac{ 27 E_0^2 - 4m^2}{ 27 E_0^2 + 4m^2}.
\end{equation}
Thus, similar pair production yields for the Bethe-Heitler and Breit-Wheeler processes can only be achieved if the Lorentz-factor $\gamma_n$ for the Bethe-Heitler nucleus is larger by a factor of $\approx 1.3$ than $\gamma_e$ for the incident electron beam in the Breit-Wheeler pair production.

To achieve a good agreement between the rates \eqref{eq: RBHcirc_under} and \eqref{eq: RBWcirc_under_av}, the prefactor of the pair production rates should also be similar. This can be achieved through a suitable choice for the target thickness $\ell$ and the nuclear charge $Z$. The Bethe-Heitler rate depends on the field through $\sim F^{5/2}$, while the averaged Breit-Wheeler rate shows a slightly different dependence of $\sim F^2$. Consequently, the resulting suitable value for the target thickness $\ell$ still depends on the considered laser field strength through
\begin{equation}
	\label{eq: l_undercrit_circ}
	\ell_{\rm circ}^{\rm (uc)}= \frac{1}{\gamma_n} \frac{\alpha Z^2}{\sqrt{2 \pi}} \left(\frac{E_0}{m}\right)^{3/2} \left(\frac{F}{F_{\rm cr}}\right)^{1/2}.
\end{equation}
When Eq.~\eqref{eq: gamma_under} is used, this formula can also be written as $\ell_{\rm circ}^{\rm (uc)} \approx \frac{ \alpha Z^2}{1.3 \, \sqrt{2 \pi}} \left[ \gamma_e \frac{F}{F_c} \right]^{1/2}$.

\subsubsection{Linearly polarized laser field}
\label{subsubsec: undercritical_lin}
Comparable rates for Bethe-Heitler and Breit-Wheeler pair production can also be achieved for a linearly polarized laser field. For field strengths $F' \ll F_{\rm cr}$, the Bethe-Heitler pair production rate in a linearly polarized laser field reads \cite{Ritus-Review}
\begin{equation}
	\label{eq: RBHlin_under}
	\mathcal{R}_{\rm BH,lin}^{\rm (uc)} = \frac{1}{\gamma_n} \frac{(Z \alpha)^2}{2 \sqrt{\pi}} m \left(\frac{\eta}{2 \sqrt{3}}\right)^{3} \exp \left(- \frac{2 \sqrt{3}}{\eta}\right).
\end{equation}
We want to compare this rate to the pair production rate of the Breit-Wheeler process in a linearly polarized laser field, which is given by \cite{Ritus-Review}
\begin{equation}
	\label{eq: RBWlin_under}
	\mathcal{R}_{\rm BW,lin}^{\rm (uc)} = \frac{3^2}{2^{11/2}} \frac{\alpha m^2}{\sqrt{\pi} \, \omega'}  \kappa^{3/2} \exp \left( - \frac{8}{3 \kappa} \right).
\end{equation}
When averaged over the bremsstrahlung spectrum, this rate becomes
\begin{equation}
	\label{eq: RBWlin_under_av1}
	\begin{split}
		& \overline{\mathcal{R}}_{\rm BW,lin}^{\rm (uc)} =\frac{\alpha m^2}{60 \sqrt{6 \pi}} \frac{\ell}{E_0} \\
		& \ \times \Bigg( - \sqrt{\pi} \text{erfc} \left(\sqrt{\frac{8}{3 \kappa_0}}\right) \left[45 \kappa_0 + 80+\frac{64}{\kappa_0} \right] \\
		& \ \ \ - \text{e }^{- \frac{8}{3 \kappa_0}}\left[\frac{147}{8} \sqrt{\frac{3}{2}} \kappa_0^{3/2} + 17 \sqrt{6 \kappa_0} + 16 \sqrt{\frac{6}{\kappa_0}} \right] \Bigg)
	\end{split}
\end{equation}
with the complementary error function $\text{erfc}(x) = 1- \text{erf}(x)$. Again, as $\kappa_0 \ll 1$, the expression in \eqref{eq: RBWlin_under_av1} can be expanded for small values of $\kappa_0$, leading to
\begin{equation}
	\label{eq: RBWlin_under_av2}
	\begin{split}
		& \overline{\mathcal{R}}_{\rm BW,lin}^{\rm (uc)} \approx \frac{3^3}{2^6} \frac{\alpha}{\sqrt{\pi}} \frac{\ell E_0^{3/2}}{\sqrt{m}} \left(\frac{F}{F_{\rm cr}}\right)^{5/2} \exp \left(- \frac{8}{3 \kappa_0} \right).
	\end{split}
\end{equation}
Therefore, to achieve similar pair production yields for the Bethe-Heitler process in \eqref{eq: RBHlin_under} and the Breit-Wheeler process in \eqref{eq: RBWlin_under_av2}, the same Lorentz-factor as for circular polarization \eqref{eq: gamma_under} is necessary. Again, we also need a field-dependent target thickness $\ell$ which results as
\begin{equation}
	\label{eq: l_undercrit_lin}
	\ell_{\rm lin}^{\rm (uc)}= \frac{1}{\gamma_n} \frac{\alpha Z^2}{2} \left(\frac{E_0}{m}\right)^{3/2} \left(\frac{F}{F_{\rm cr}}\right)^{1/2}.
\end{equation}

\subsection{Nonperturbative pair production in the overcritical regime}
\label{subsec: overcritical}

\subsubsection{Circularly polarized laser field}
\label{subsubsec: overcritical_circ}
So far, we have considered the comparison of the Bethe-Heitler and Breit-Wheeler rate in the undercritical regime, where $F' \ll F_{\rm cr}$. In the following, we compare both processes in the complementary regime of overcritical field strengths $F' \gg F_{\rm cr}$. For a circularly polarized laser field, the rate of the Bethe-Heitler pair production process reads \cite{Milstein} 
\begin{equation}
	\label{eq: RBHcirc_over}
	\mathcal{R}_{\rm BH,circ}^{\rm (oc)} = \frac{1}{\gamma_n} \frac{13}{6 \sqrt{3} \pi} \left(Z \alpha \right)^2 m \eta \left[\ln \left(\frac{\eta}{2 \sqrt{3}}\right) - C - \frac{58}{39} \right].
\end{equation}
Here, $C \approx 0.577$ denotes the Euler-Mascheroni constant. \\
The Breit-Wheeler rate in the overcritical regime for a fixed frequency $\omega '$ is given by \cite{Ritus-Review}
\begin{equation}
	\label{eq: RBWcirc_over}
	\mathcal{R}_{\rm BW,circ}^{\rm (oc)} = \frac{15}{14} \frac{\Gamma^4(2/3)}{\pi^2} \frac{\alpha m^2}{\omega'} \left(3 \kappa \right)^{2/3},
\end{equation}
with $\kappa = \frac{2 \omega'}{m} \frac{F}{F_{\rm cr}} \gg 1$. Note, that the expression for the Breit-Wheeler rate in Eq.~\eqref{eq: RBWcirc_over} can only be used for $\kappa \ll \frac{1}{\alpha^{3/2}} \approx 1600$ \cite{Ritus-Review, Narozhny}, putting an upper limit on the considered field strength. In the following, we will consequently only consider fields which fulfill this condition. When averaging this rate over the bremsstrahlung spectrum, energies $\omega'$ ranging from $0$ to the electron beam energy $E_0$ need to be taken into account. However, if the energy $\omega'$ is too small, the pair production process takes place in the undercritical regime of Sec.~\ref{subsec: undercritical}. Only for sufficiently large values of $\omega'$, the overcritical regime is reached. Therefore, we split the calculation of the averaged Breit-Wheeler rate into a region with small field strengths $F' < F_{\rm cr}$, where Eq.~\eqref{eq: RBWcirc_under} holds, and a region of larger field strengths $F' > F_{\rm cr}$, where the rate of the overcritical regime in Eq.~\eqref{eq: RBWcirc_over} is taken into account. $\overline{\mathcal{R}}_{\rm BW,circ}^{\rm (oc)}$ consequently results from 
\begin{equation}
	\label{eq: RBWcirc_over_av1}
	\begin{split}
		\overline{\mathcal{R}}_{\rm BW,circ}^{\rm (oc)} = & \int_0^{\frac{m}{2} \frac{F_{\rm cr}}{F}} I_{\gamma}(\omega' , \ell) \mathcal{R}_{\rm BW,circ}^{\rm (uc)} (\omega') \, \dd \omega' \\
		& + \int_{\frac{m}{2} \frac{F_{\rm cr}}{F}}^{E_0} I_{\gamma}(\omega', \ell) \mathcal{R}_{\rm BW,circ}^{\rm (oc)} (\omega ') \, \dd \omega'.
	\end{split}
\end{equation}
As the second integration of \eqref{eq: RBWcirc_over_av1} contributes the most significantly to the averaged rate $\overline{\mathcal{R}}_{\rm BW,circ}^{\rm (oc)}$, the integration over the tunneling rate will be disregarded in the following. The relevant integral can be calculated analytically, yielding
\begin{equation}
	\label{eq: RBWcirc_over_av2}
	\begin{split}
		& \overline{\mathcal{R}}_{\rm BW,circ}^{\rm (oc)} = \frac{5 \cdot 3^{5/3}}{7 \pi^2} \Gamma^4\left(\frac{2}{3} \right) \frac{\alpha m^2}{E_0} \, \ell  \kappa_0^{2/3}\\
		& \ \ \ \times \left( - \frac{27}{5} + 4 \kappa_0^{1/3} + \frac{2}{\kappa_0^{2/3}} - \frac{3}{5 \kappa_0^{5/3}} \right)
	\end{split}
\end{equation}
with $\kappa_0 = \frac{2 E_0}{m} \frac{F}{F_{\rm cr}}$. In the overcritical regime $\kappa_0 \gg 1$, such that the first two terms of the sum in \eqref{eq: RBWcirc_over_av2} are dominating while the last two can be neglected. We can therefore approximately write the averaged Breit-Wheeler rate in the overcritical regime as
\begin{equation}
	\label{eq: RBWcirc_over_av3}
	\begin{split}
		\overline{\mathcal{R}}_{\rm BW,circ}^{\rm (oc)} & \approx \frac{10 \times 3^{5/3}}{7 \pi^2} \Gamma^4\left(\frac{2}{3} \right) \frac{\alpha m^2}{E_0} \, \ell \kappa_0  \left[ 1 - \frac{27}{20 \kappa_0^{1/3}} \right].
	\end{split}
\end{equation}
The averaged Breit-Wheeler rate depends on a term linear to $\kappa_0$ multiplied with a more complicated dependence written in brackets of Eq.~\eqref{eq: RBWcirc_over_av3}. The Bethe-Heitler rate in Eq.~\eqref{eq: RBHcirc_over} depends on $\eta$ in a similar manner. To achieve the same pair production rate for both processes, it is therefore reasonable to set $\kappa_0 = \eta$ and to consequently use the same Lorentz factor $\gamma_n = \gamma_e = \frac{E_0}{m}$.

Further, the required target thickness for the creation of bremsstrahlung is obtained by equating the resulting pair production rates and results as
\begin{equation}
	\label{eq: l_overcrit_circ}
	\ell_{\rm circ}^{\rm (oc)}= \frac{91}{540 \times 3^{1/6} } \frac{ \pi}{\Gamma^4 (2/3)} Z^2 \alpha \frac{\text{ln} \left( \frac{\eta}{2 \sqrt{3}} \right) - C - \frac{58}{39}}{1 - \frac{27}{20 \eta^{1/3}}}
\end{equation}
with $\eta = \left( 1 + \beta_n \right) \gamma_n \frac{F}{F_{\rm cr}} = \frac{2 E_0}{m} \frac{F}{F_{\rm cr}}$.

\subsubsection{Linearly polarized laser field}
\label{subsubsec: overcritical_lin}

We can further compare the Bethe-Heitler and Breit-Wheeler rates for a linearly polarized field. The Bethe-Heitler rate for an elliptically polarized laser wave was obtained up to logarithmic accuracy in Eq.~(38) of \cite{Milstein}. In order to provide a fair comparison with the more accurate Eq.~\eqref{eq: RBHcirc_over} for circular polarization, we have followed the calculation of Ref.~\cite{Milstein} and derived the beyond-logarithmic corrections for the Bethe-Heitler rate in a linearly polarized field. Our derivation, which is given in Appendix~ \ref{appendix}, leads to the result
\begin{equation}
	\label{eq: RBHlin_over}
	\begin{split}
	\mathcal{R}_{\rm BH,lin}^{\rm (oc)} &= \frac{1}{\gamma_n} \frac{13}{3 \sqrt{3} \pi^2} \left(Z \alpha \right)^2 m \eta \\
	& \times \left[  \text{ln} \left( \frac{\eta}{2 \sqrt{3}} \right) - C - \frac{141}{65} + \frac{7}{13} \text{ln} \left(2\right)\right].
	\end{split}
\end{equation}
Up to logarithmic accuracy, this expression coincides with Eq.~(38) in \cite{Milstein} when the latter is applied to linear polarization.

On the other hand, the Breit-Wheeler rate with a fixed frequency $\omega'$ is given by \cite{Ritus-Review}
\begin{equation}
	\label{eq: RBWlin_over}
	\mathcal{R}_{\rm BW,lin}^{\rm (oc)} = \frac{135}{56} \frac{\Gamma^7\left( 2/3 \right)}{ \pi^4} \frac{\alpha m^2}{\omega'} \left( \frac{3 \kappa}{2} \right)^{2/3}.
\end{equation}
The averaging according to Eq.~\eqref{eq: RBWcirc_over_av1} approximately gives
\begin{equation}
	\label{eq: RBWlin_over_av}
	\begin{split}
		\overline{\mathcal{R}}_{\rm BW,lin}^{\rm (oc)} & \approx \frac{135}{14 \, \pi^4} \left(\frac{3}{2}\right)^{2/3} \Gamma^7\left(\frac{2}{3} \right) \frac{\alpha m^2}{E_0} \, \ell \kappa_0 \\
		& \times \left[ 1 - \frac{27}{20 \kappa_0^{1/3}} \right].
	\end{split}
\end{equation}
To achieve the same pair production rate for both the Bethe-Heitler and Breit-Wheeler process, we again choose the same Lorentz-factor $\gamma_n= \frac{E_0}{m}$ and a target thickness according to

\begin{equation}
\label{eq: l_overcrit_lin}
\begin{split}
\ell_{\rm lin}^{\rm (oc)} = & \frac{182}{405 \sqrt{3}} \left(\frac{2}{3}\right)^{2/3} \frac{ \pi^2}{\Gamma^7(2/3)} Z^2 \alpha \\
& \times \frac{\text{ln} \left( \frac{ \eta}{2 \sqrt{3}} \right) - C - \frac{141}{65} + \frac{7}{13} \text{ln} \left(2\right)}{1 - \frac{27}{20 \eta^{1/3}}}.
\end{split}
\end{equation}

\section{Results and Discussion}
\label{sec: results}

In the following, we will compare numerical results for the pair production rates of the Bethe-Heitler and bremsstrahlung-driven Breit-Wheeler processes in the undercritical ($F'\ll F_{\rm cr}$) and the overcritical ($F'\gg F_{\rm cr}$) interaction regimes. Also the corresponding spectra of bremsstrahlung photons versus equivalent photons shall be compared.

In the previous Sec.~\ref{sec: comparison} we have obtained expressions for the nuclear Lorentz-factor $\gamma_n$ and the bremsstrahlung converter thickness $\ell$, such that identical pair production rates for the nonlinear Bethe-Heitler and Breit-Wheeler process are achieved. For the required ratio $\gamma_n/\gamma_e$, a constant value of about 1.3 was found, whereas the target thickness shows a rather complicated parameter dependence, in particular on the applied laser field strength $F$. In a Breit-Wheeler experiment it would in principle be conceivable to vary the target thickness with the laser amplitude $F$ by utilizing a suitable number of different converter foils. In this scenario, the magnitude of each measured Breit-Wheeler rate would correspond exactly to an associated Bethe-Heitler rate with properly chosen nuclear Lorentz factor. However, in view of the experimental feasibility, it appears more convenient to perform the measurements with a uniform foil thickness. Therefore, we will assume a fixed value of the converter foil thickness $\ell$ in each interaction regime, and for each considered laser polarization. 

Another -- even more important -- reason for taking a fixed target thickness is that intense laser fields in experiment have the form of finite pulses, whereas the nonlinear Bethe-Heitler and Breit-Wheeler rates in Sec.~\ref{sec: comparison} assume infinitely extended laser waves of given amplitude $F$. Instead, laser pulses have an envelope $f(\varphi)$, where $\varphi$ denotes the laser phase, so that the laser amplitude varies as $F = F_0 f(\varphi)$. When the pulse is relatively long and the amplitude variation slow, one can apply the locally monochromatic approximation (LMA) combined with the slowly varying envelope approximation (SVEA) to obtain the pair yield via an expression of the form $\int \mathcal{R}\big(F_0 f(\varphi)\big)\,{\rm d}\varphi$ \cite{Review4, Heinzl2020, King2021,Tang-Spin, Eckey2022, King2024, Seipt2025}. Thus, various field amplitudes are probed and contribute to the pair production. Therefore, it is relevant to see how closely the Bethe-Heitler rate $\mathcal{R}_{\rm BH}(F)$ and Breit-Wheeler rate $\overline{\mathcal{R}}_{\rm BW}(F)$ agree in a whole range of field amplitudes, when the converter thickness is kept constant. 

The envisaged parameters of the upcoming Breit-Wheeler experiment at the CALA facility \cite{CALA} serve us as an orientation, which relies on electron beams of $E_0=2.5$\,GeV energy to generate bremsstrahlung in a 50\,$\mu$m thick tungsten foil ($L_{\rm rad} \approx 3.3$\,mm, corresponding to $\ell\approx 1.5 \times 10^{-2}$) and 30\,fs laser pulses with peak intensities of $\sim 10^{21}$-$10^{22}$\,W/cm$^2$, corresponding to field strengths on the order of $F\sim 10^{-4} F_{\rm cr}$. We note that $\kappa_0\approx 1$ for these parameters, placing the CALA experiment in between the undercritical and overcritical field regimes. For somewhat smaller field strengths, the pair production will proceed as a tunneling process (Sec.~\ref{subsec: undercritical_results}), while for substantially higher field strengths the over-barrier regime is entered (Sec.~\ref{subsec: overcritical_results}). 

\subsection{Nonperturbative pair production in the undercritical regime}
\label{subsec: undercritical_results}

\subsubsection{Comparison of the pair production rates}
\label{subsubsec: undercritical_rates}

\begin{figure}
	\centering
	\vspace{-0.25cm}
	\includegraphics[width=0.48\textwidth]{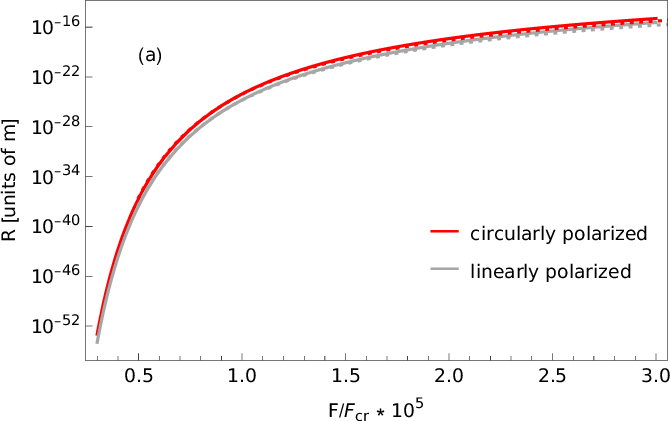}
	\vspace{0.15cm}
	\includegraphics[width=0.48\textwidth]{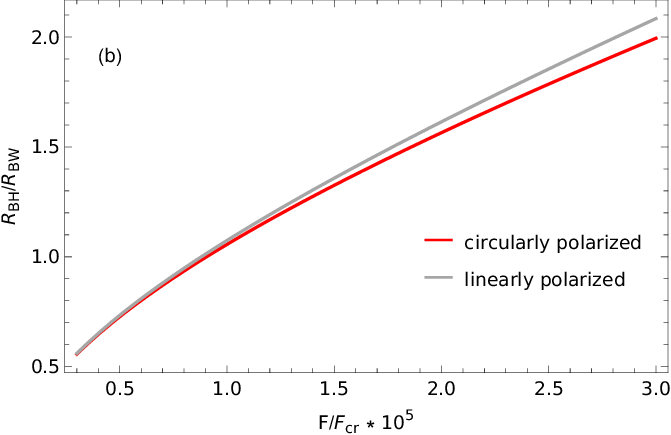}
	\vspace{-0.5cm}
	\caption[justification=justified]{Panel a) shows the Bethe-Heitler (solid lines) and Breit-Wheeler (dashed lines) rates with the Lorentz-factor from Eq.~\eqref{eq: gamma_under} and target thicknesses $\ell$ for a circularly polarized field from Eq.~\eqref{eq: l_undercrit_circ} and for a linearly polarized laser field from Eq.~\eqref{eq: l_undercrit_lin}. The results for a circularly polarized field are shown in red while the ones for linear polarization are depicted in gray. A nuclear charge $Z=1$ and an electron beam energy of $E_0= 2.5 \, {\rm GeV}$ are considered while the target thicknesses are evaluated at $F=10^{-5} \, F_{\rm cr}$ and result as $\ell_{\rm circ}^{\rm (uc)} = 0.50 \times 10^{-3}$ and $\ell_{\rm lin}^{\rm (uc)} = 0.62 \times 10^{-3}$. The rates are shown as a function of the field $F$ in units of $10^{-5} \, F_{\rm cr}$. In panel b) the ratio of the resulting Bethe-Heitler rate with the Breit-Wheeler rate is depicted.}
	\label{fig: undercritical_E0}
\end{figure}

In Sec.~\ref{subsec: undercritical}, we have found appropriate values for the Lorentz-factor in Eq.~\eqref{eq: gamma_under} and for the target thickness $\ell$ for a circularly \eqref{eq: l_undercrit_circ} as well as for a linearly polarized laser field \eqref{eq: l_undercrit_lin}. In Fig.~\ref{fig: undercritical_E0} the resulting pair production rates are depicted as a function of the applied field strengths $F$ in units of $F_{\rm cr}$. The top panel of Fig.~\ref{fig: undercritical_E0} shows the pair production rates for both linear (gray) and circular (red) laser field polarization for a nuclear charge $Z=1$ and an electron beam energy $E_0 = 2.5 \, {\rm GeV}$, corresponding to $\gamma_e \approx 5 \times 10^{3}$. Further, $F=10^{-5} F_{\rm cr}$ is used as a reference field strength for the target thickness such that thicknesses of $\ell_{\rm circ}^{\rm (uc)} = 0.50 \times 10^{-3}$ for circular and $\ell_{\rm lin}^{\rm (uc)} = 0.62 \times 10^{-3}$ for linear polarization are considered. The solid lines show the pair production rate of the Bethe-Heitler process while the dashed lines depict the Breit-Wheeler rate. In the second panel of Fig.~\ref{fig: undercritical_E0} the ratio of the considered Bethe-Heitler and Breit-Wheeler rates is displayed.

Fig.~\ref{fig: undercritical_E0} shows that the choice for $\gamma_n$ and $\ell$ lead to a good agreement of the Bethe-Heitler and Breit-Wheeler rates in the range $0.04 \lesssim \eta \lesssim 0.4$ and $0.03 \lesssim \kappa_0 \lesssim 0.3$ respectively. As the target thickness is considered for $F=10^{-5} \, F_{\rm cr}$, the rates exactly coincide for the reference field strength. Nevertheless, even though the target thickness was evaluated at $F = 10^{-5} F_{\rm cr}$, the ratio of both rates demonstrates that a good agreement between the rates can still be obtained in a wider range of field strengths, as for example $\mathcal{R}_{\rm BH}^{(\rm uc)} / \overline{\mathcal{R}}_{\rm BW}^{\rm (uc)} \approx 0.75$ for $F = 0.5 \times 10^{-5} F_{\rm cr}$. As a consequence, in the logarithmic representation of panel a) the solid Bethe-Heitler and dashed Breit-Wheeler rates lie practically on top of each other for a given laser polarization. Our results imply moreover that, if a multi-cycle laser pulse of maximum amplitude $F_0 = 10^{-5}F_{\rm cr}$ and slowly varying envelope is applied in experiment, the ratio between the Bethe-Heitler and the bremsstrahlung-averaged Breit-Wheeler pair yields will still be close to 1 for the chosen target thickness.

The previous consideration has shown that the Lorentz-factor of the nucleus in the Bethe-Heitler process needs to be $\approx 1.3$ times larger than the one of the electron beam which is used to generate the bremsstrahlung photons for the Breit-Wheeler process. This immediately results from the different exponential functions of both processes, as the numerical prefactor in the Breit-Wheeler exponential stays unchanged after taking the average over the bremsstrahlung spectrum.

A physically intuitive argument allows us to understand why -- in the undercritical  regime -- the Bethe-Heitler process by a nucleus with Lorentz factor $\gamma_n$ is less effective than the Breit-Wheeler process by a photon with energy $\omega' = \gamma_n m$. To this end, let us compare both processes in another frame of reference which is reached by a Lorentz boost with Lorentz factor $\gamma_n$. In the Bethe-Heitler case, this coincides with the frame where the nucleus is at rest and the laser photons have a largely increased energy of $\omega_{\rm rf} = (1 + \beta_n) \gamma_n \omega$. For example, when $\omega = 1$\,eV in the laboratory frame and $\gamma_n =  5000$, one obtains $\omega_{\rm rf} \approx 10$\,keV. When the same Lorentz boost is applied to the Breit-Wheeler scenario, the laser photons attain the upshifted energy $\omega_{\rm rf}$ as well, whereas the energy of the counterpropagating non-laser photon is reduced to $\omega'_{\rm rf} = (1- \beta_n) \gamma_n$. For $\omega' = \gamma_n m = 2.5$\,GeV, the boosted frequency results as $\omega'_{\rm rf} \approx 250$\,keV. Comparing these two scenarios we see that, while the boosted laser field is the same in each case, the Bethe-Heitler process involves a nucleus at rest whose static Coulomb field has no time dependence. In contrast, the field of the boosted $\gamma$-photon in the Breit-Wheeler process is oscillatory and thus time-dependent. Even though the value of $\omega_{\rm rf}'$ in our example is rather small, it can still play a significant role in the tunneling regime, as the exponential rate is very sensitive to small changes in the supplied energy. This reasoning can qualitatively explain why the Breit-Wheeler process in the undercritical regime at $\omega' = \gamma_n m$ yields larger pair production rates than the corresponding Bethe-Heitler process by a nucleus with Lorentz factor $\gamma_n$.

\subsubsection{Comparison of the spectra for bremsstrahlung and equivalent photons}
\label{subsubsec: undercritical_spectra}
In Sec.~\ref{subsubsec: undercritical_circ} and \ref{subsubsec: undercritical_lin}, a suitable choice of the Lorentz-factor $\gamma_n$ for the Bethe-Heitler pair production process and the target thickness $\ell$ of the bremsstrahlung for the Breit-Wheeler pair production has been shown to successfully lead to similar pair production yields. Further, both the spectrum of the equivalent photons in the Bethe-Heitler process and the bremsstrahlung spectrum in the Breit-Wheeler process show an approximate dependence on the energy through $\sim \frac{1}{\omega'}$. Therefore, for the chosen values for $\gamma_n$ in \eqref{eq: gamma_under} and the target thickness $\ell$ in \eqref{eq: l_undercrit_circ} and \eqref{eq: l_undercrit_lin}, also the spectra \eqref{eq: spectrum equivalent photon_exact} and \eqref{eq: spectrum bremsstrahlung} are expected to show a similar dependence on the energy $\omega'$.

\begin{figure}[t]  
	\vspace{-0.25cm}
	\begin{center}
		\includegraphics[width=0.48\textwidth]{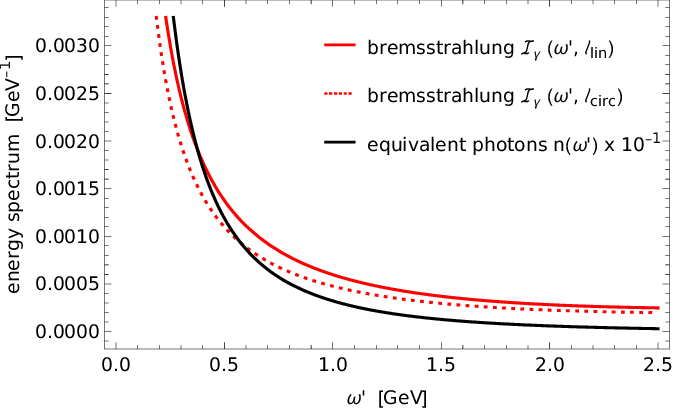}
	\end{center}
	\vspace{-0.5cm} 
	\caption[justification=justified]{Comparison of the different photon energy spectra. The spectrum for the equivalent photons \eqref{eq: spectrum equivalent photon_exact} multiplied by a factor $10^{-1}$ is shown through the solid black line. The spectrum for the bremsstrahlung \eqref{eq: spectrum bremsstrahlung} with the target thickness $\ell_{\rm lin}^{\rm (uc)} = 0.62 \times 10^{-3}$ for a linearly polarized laser field \eqref{eq: l_undercrit_lin} is shown in the solid red line and the dashed red line shows the corresponding spectrum for the target thickness $\ell_{\rm circ}^{\rm (uc)} = 0.50 \times 10^{-3}$ needed for a circularly polarized laser field \eqref{eq: l_undercrit_circ}. A nuclear charge $Z=1$ and an electron beam energy $E_0 = \gamma_e m = 2.5 \, \text{GeV}$ are considered, while the nuclear Lorentz factor is $\gamma_n \approx 1.3 \, \gamma_e$.}
	\label{fig: undercritical_spectra}
\end{figure}

Fig.~\ref{fig: undercritical_spectra} shows the comparison of the resulting spectra of the equivalent photons up to $\omega' = E_0$ with the bremsstrahlung spectra for both a linear (solid red line) and circular (dashed red line) polarized laser field. It shows that both spectra for bremsstrahlung, even though they do not exactly coincide, show the same energy dependence. The equivalent photon spectrum -- multiplied with a factor $10^{-1}$ -- is shown by the solid black line in Fig.~\ref{fig: undercritical_spectra}. While the course of the spectrum shows similarities with both bremsstrahlung spectra, they deviate by approximately one order of magnitude. Even though the similar pair production rates suggest a resemblance between the equivalent photon and bremsstrahlung spectrum, Fig.~\ref{fig: undercritical_spectra} shows that this is not the case for the undercritical regime. Note however, that mostly large values for $\omega'$ contribute to the average of the Breit-Wheeler rate over the bremsstrahlung spectrum as here the Breit-Wheeler rate is largest \cite{Golub2022}. Towards the bremsstrahlung endpoint energy, the spectra only show a relatively small difference. For example for $\omega' = E_0 = 2.5 \,$GeV, the ratio between both spectra results as $I_{\gamma} (\omega') / n (\omega') \approx 0.35, \ 0.49 \ \text{and} \ 0.60$ for $F= 0.5 \times 10^{-5} \, F_{\rm cr}, \ 10^{-5} \, F_{\rm cr} \ \text{and} \ 1.5 \times 10^{-5} \, F_{\rm cr}$, respectively.

Next, we analyze the applicability of the equivalent photon method (or Weizsäcker-Williams approximation) to quantitatively describe the Bethe-Heitler process as a Breit-Wheeler process averaged over the equivalent photon spectrum, according to $\mathcal{R}_{\rm BH} \approx \mathcal{R}_{\rm BW}^{\rm (eq)}$ with

\begin{equation}
	\label{eq: BH as BW}
	\mathcal{R}_{\rm BW}^{\rm (eq)} := \int \mathcal{R}_{\rm BW} (\omega') \, n(\omega') \dd \omega'.
\end{equation}

This equivalent photon method is known to work well for the pair production processes in which only one photon is involved (i.e. \eqref{BW} and \eqref{BH} with $n=1$) \cite{Landau, STAR}. However, for the nonlinear Bethe-Heitler process in the undercritical field regime, the equivalent photon method turns out to vastly overestimate the rate. As previously discussed, the numerical factors in the exponential functions of the Bethe-Heitler and Breit-Wheeler rates are different. An adjustment of the Lorentz factors was necessary in order to obtain coinciding pair production rates. Further, the energy $\omega'$ in the bremsstrahlung spectrum is limited up to $\gamma_e  m$, while larger energies can contribute in the equivalent photon spectrum. In fact, the main contribution to the integral in Eq.~\eqref{eq: BH as BW} results from $\omega' \gg \gamma_e m$, leading to a larger averaged rate and consequently a big difference to the actual Bethe-Heitler pair production rate. Consequently, for the undercritical regime, the ratio of the actual Bethe-Heitler rate and the averaged Breit-Wheeler rate over the equivalent photon spectrum is strongly field dependent and for example yields $\mathcal{R}_{\rm BH} / \mathcal{R}_{\rm BW}^{\rm (eq)} \approx 10^{-11}$ for $F/F_{\rm cr} = 10^{-5}$, where $n(\omega')$ is taken from Eq.~\eqref{eq: spectrum equivalent photon_exact}. Evidently, in this case the idea to average the Breit-Wheeler rate over the equivalent photon spectrum to obtain the Bethe-Heitler rate, does not apply. 

\begin{figure}[t]  
	\vspace{-0.25cm}
	\begin{center}
		\includegraphics[width=0.48\textwidth]{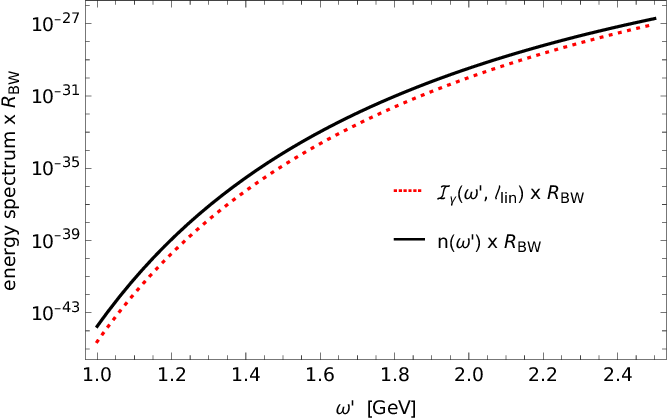}
	\end{center}
	\vspace{-0.5cm} 
	\caption[justification=justified]{Comparison of the integrands obtained by multiplying the Breit–Wheeler rate for linear polarization with the bremsstrahlung spectrum (dashed red line) and the equivalent photon spectrum (solid black line), shown as functions of the energy $\omega'$. A field strength of $F = 10^{-5} \,F_{\rm cr}$, $E_0 = 2.5$ GeV and $\gamma_n \approx 1.3 \, E_0 / m$ are considered.}
	\label{fig: undercritical_integrand}
\end{figure}

However, by comparison with the previously discussed average over the bremsstrahlung spectrum we can introduce a modification to the Weizsäcker-Williams method which allows to approximately obtain the Bethe-Heitler rate from an average of the Breit-Wheeler rate over the equivalent photon spectrum. To this end, the integral in Eq.~\eqref{eq: BH as BW} is restricted by hand to an upper limit of $\gamma_e m$, where $\gamma_e \approx \gamma_n / 1.3$. The corresponding integrand for $F=10^{-5} F_{\rm cr}$ and $\gamma_n \approx 1.3 \times 2.5 \,{\rm GeV} / m$ is depicted by the solid black line in Fig.~\ref{fig: undercritical_integrand}. It is compared to the previously considered integrand in Eq.~\eqref{eq: RBWcirc_under_av1} of the bremsstrahlung spectrum multiplied with the Breit-Wheeler rate (dashed red line). The modified Weizs\"acker-Williams method yields $\mathcal{R}_{\text{BH}} / \mathcal{R}_{\text{BW}}^{\text{(eq)}} \approx 0.34, \ 0.49 \ \text{and} \ 0.59 $ for $F= 0.5 \times 10^{-5} \, F_{\rm cr}, \ 10^{-5} \, F_{\rm cr}, \ \text{and} \ 1.5 \times 10^{-5} \, F_{\rm cr}$ respectively, which represents a fairly good agreement. We note that this rate ratio closely coincides with the ratio of the bremsstrahlung and equivalent photon spectrum at $\omega' = E_0$.

By introducing the upper limit, the integration in Eq.~\eqref{eq: BH as BW} can be evaluated analytically by using the equivalent photon spectrum given in Eq.~\eqref{eq: spectrum equivalent photon}. The resulting $\mathcal{R}_{\rm BW}^{\rm (eq)}$ is proportional to an exponential function of the usual Breit-Wheeler form $\exp \left(-\frac{8}{3 \kappa_e}\right)$, where $\kappa_e = 2 \gamma_e \frac{F}{F_{\rm cr}}$. Accordingly, the previously introduced upper limit $\gamma_e m \approx \gamma_n m / 1.3$ for the integration in the Weizs\"acker-Williams method ensures that the exponential of the Breit-Wheeler rate goes over into the Bethe-Heitler exponential $\exp \left(-\frac{2 \sqrt{3}}{ \eta}\right)$ with $\eta \approx 2 \gamma_n \frac{F}{F_{\rm cr}}$.

\subsection{Nonperturbative pair production in the overcritical regime}
\label{subsec: overcritical_results}

\subsubsection{Comparison of the pair production rates}
\label{subsubsec: overcritical_rates}
\begin{figure}
	\centering
	\vspace{-0.25cm}
	\includegraphics[width=0.48\textwidth]{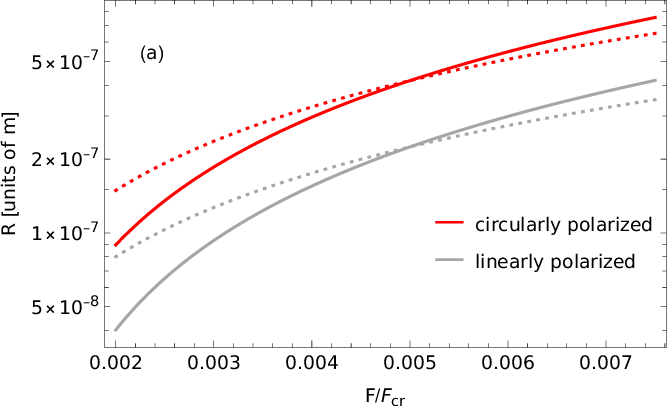}
	\vspace{0.15cm}
	\includegraphics[width=0.48\textwidth]{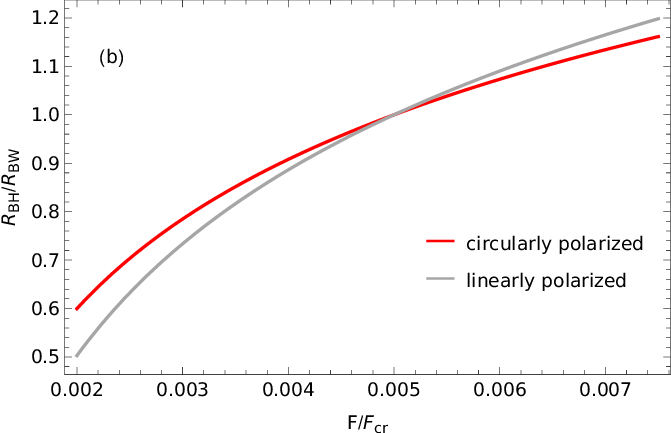}
	\vspace{-0.5cm}
	\caption[justification=justified]{Panel a) shows the Bethe-Heitler (solid lines) and Breit-Wheeler (dashed lines) rates with the Lorentz-factor $\gamma_n = \gamma_e = \frac{E_0}{m}$ and target thicknesses $\ell$ resulting from Eqs.~\eqref{eq: l_overcrit_circ} and \eqref{eq: l_overcrit_lin}. The results for a circularly polarized field are shown in red while the ones for linear polarization are depicted in gray. A nuclear charge $Z=1$ and an electron beam energy of $E_0= 10 \, {\rm GeV}$ are considered while the target thicknesses are evaluated at a reference field strength $F=0.005 \, F_{\rm cr}$ and result as $\ell_{\rm circ}^{\rm (oc)} = 0.0025$ and $\ell_{\rm lin}^{\rm (oc)} = 0.0037$. The rates are shown as a function of the field $F$ in units of the critical field $F_{\rm cr}$. In panel b) the ratio of the resulting Bethe-Heitler rate with the Breit-Wheeler rate is depicted.}
	\label{fig: overcritical_Z}
\end{figure}

In Sec.~\ref{subsec: overcritical}, we have found that, in order to achieve similar pair production rates in the overcritical regime ($F' \gg F_{\rm cr}$), the Lorentz-factors of the Bethe-Heitler and Breit-Wheeler process can be chosen equally. Additionally, Eqs.~\eqref{eq: l_overcrit_circ} and \eqref{eq: l_overcrit_lin} provide the necessary target thickness for circular and linear polarization, respectively. The top panel of Fig.~\ref{fig: overcritical_Z} shows the resulting rates of the Bethe-Heitler and Breit-Wheeler process for a nuclear charge $Z=1$ and an initial electron beam energy of $E_0 = 10 \, \text{GeV}$, corresponding to $\gamma_e=\gamma_n \approx 2 \times 10^4$. For the target thickness $\ell$, a reference field strength of $F = 0.005 \, F_{\rm cr}$ is used, which ensures that the overcritical regime with $F' \gg F_{\rm cr}$ is reached. The resulting target thicknesses at $F = 0.005 \, F_{\rm cr}$ are given by $\ell_{\rm circ}^{\rm (oc)} = 0.0025$ and $\ell_{\rm lin}^{\rm (oc)} = 0.0037$. The Bethe-Heitler pair production rate is shown by solid lines, while the Breit-Wheeler rate is depicted in dashed lines. Further, the results for circular polarization are shown in red and for linear polarization in gray.

Fig.~\ref{fig: overcritical_Z} shows that with our choice for the Lorentz-factor $\gamma_n$ and the target thickness $\ell$, similar pair production rates are achieved in the considered range $80 \lesssim \eta, \kappa_0 \lesssim 300$. The ratio of the Bethe-Heitler and Breit-Wheeler rate (depicted in the second panel of Fig.~\ref{fig: overcritical_Z}) lies close to $1$ for a wide range of field strengths.

\subsubsection{Comparison of the spectra for bremsstrahlung and equivalent photons}
\label{subsubsec: overcritical_spectra}
\begin{figure}[t]  
	\vspace{-0.25cm}
	\begin{center}
		\includegraphics[width=0.48\textwidth]{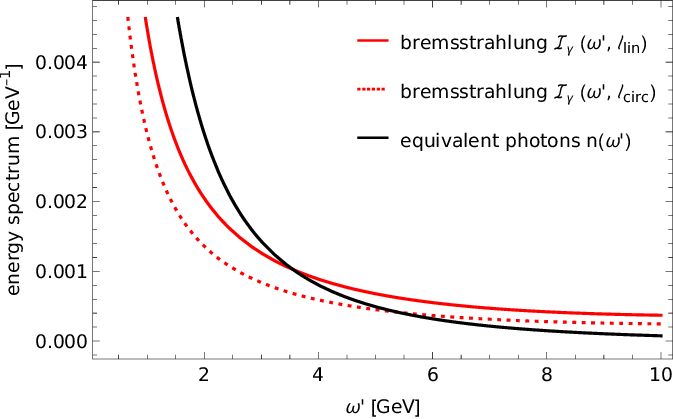}
	\end{center}
	\vspace{-0.5cm} 
	\caption[justification=justified]{Comparison of the different photon energy spectra. The spectrum for the equivalent photons \eqref{eq: spectrum equivalent photon_exact} is shown through the solid black line. The spectrum for the bremsstrahlung \eqref{eq: spectrum bremsstrahlung} with the target thickness $\ell_{\rm lin}^{\rm (oc)} = 0.0037$ for a linearly polarized laser field \eqref{eq: l_overcrit_lin} is shown in the solid red line and the dashed red line shows the same spectrum for the target thickness $\ell_{\rm circ}^{\rm (oc)} =0.0025$ needed for a circularly polarized laser field \eqref{eq: l_overcrit_circ}. A nuclear charge $Z=1$ and an electron beam energy $E_0 = 10 \, \text{GeV}$ are considered.}
	\label{fig: overcritical_spectra}
\end{figure}

Apart from the pair production rates, also the resulting spectra of bremsstrahlung and equivalent photons can be compared. Fig.~\ref{fig: overcritical_spectra} shows the spectra for energies $\omega'$ up to the electron-beam energy $E_0 = 10 \, \text{GeV}$. The equivalent photon spectrum is depicted by solid black lines, while the bremsstrahlung spectra for linear (circular) polarization are shown by solid (dashed) red lines.

In contrast to Fig.~\ref{fig: undercritical_spectra} for the undercritical regime, Fig.~\ref{fig: overcritical_spectra} shows that the spectrum of the equivalent photons for the Bethe-Heitler process does approximately align with the bremsstrahlung spectrum of the Breit-Wheeler process. As a result, in the overcritical regime the considered target thickness to achieve a good agreement between the pair production rates, also leads to similarities between the bremsstrahlung spectra and the equivalent photon spectrum. When comparing the spectra of the under- and overcritical regime it becomes evident that this difference directly results from the larger values for the necessary target thickness. As $I_{\gamma}(\omega', \ell)$ is linear in the target thickness $\ell$, the bremsstrahlung spectrum is increased compared to the undercritical regime, while the equivalent photon spectrum is unaffected by a different target thickness. 

\begin{figure}[t]  
	\vspace{-0.25cm}
	\begin{center}
		\includegraphics[width=0.48\textwidth]{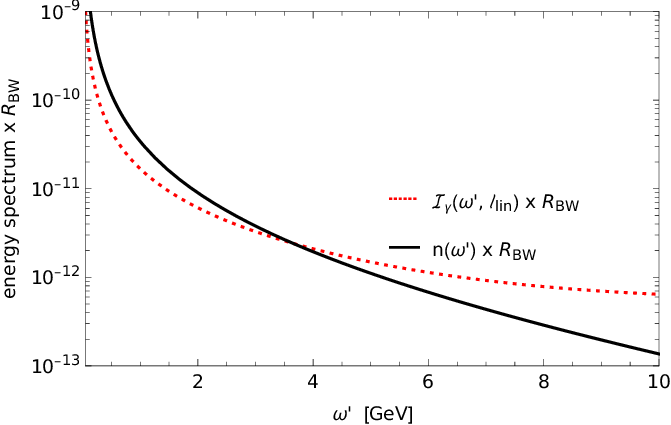}
	\end{center}
	\vspace{-0.5cm} 
	\caption[justification=justified]{Comparison of the integrands obtained by multiplying the Breit–Wheeler rate for linear polarization with the bremsstrahlung spectrum (dashed red line) and with the equivalent photon spectrum (solid black line), shown as functions of the energy $\omega'$. A field strength of $F = 0.005 \,F_{\rm cr}$, $E_0 = 10$ GeV and $\gamma_n = E_0 / m$ are considered.}
	\label{fig: overcritical_integrand}
\end{figure}

In the overcritical regime, we find that the Weizs\"acker-Williams method works well. In contrast to the undercritical regime, the main contribution to the integral of Eq.~\eqref{eq: BH as BW} results from $\omega' \ll \gamma_e m$, which facilitates the description of the Bethe-Heitler rate as a Breit-Wheeler process averaged over the equivalent photon spectrum. For example for $\gamma_n = 10 \, \text{GeV} / m$ the ratio of the actual Bethe-Heitler rate and the averaged Breit-Wheeler rate over the equivalent photon spectrum yields $\mathcal{R}_{\rm BH} / \mathcal{R}_{\rm BW}^{\rm (eq)} \approx 1.17 , \ 1.02$ and $0.97$ for $F = 0.002 \, F_{\rm cr} , \ F = 0.005 \, F_{\rm cr}$ and $F = 0.008 \, F_{\rm cr}$ respectively. Consequently, in the overcritical regime, the Bethe-Heitler pair production process can be approximately described as a Breit-Wheeler process by using the usual Weizs\"acker-Williams method. This is further corroborated by Fig.~\ref{fig: overcritical_integrand}, where the corresponding integrands of the Weizs\"acker-Williams approximation and the averaging over the bremsstrahlung spectrum are depicted, considering the overcritical regime ($\omega' > 0.0511$ GeV). Here, mainly small energies $\omega'$ contribute for which the integrands show a good resemblance. For even smaller $\omega'$ the undercritical regime is reached where the integrand strongly decreases, such that its contribution can be neglected in the integral.

\vspace{0.5cm}
\section{Conclusion}
\label{sec: conclusion}

\noindent
Nonlinear Bethe-Heitler pair production by a relativistic nucleus colliding with an intense laser field and nonlinear Breit-Wheeler pair production by bremsstrahlung $\gamma$-photons and an intense laser field have been analyzed in a comparative study. Our consideration has been motivated by the similar shapes of the equivalent photon spectrum of a relativistically moving nucleus and the energy spectrum of (real) bremsstrahlung photons from a high-energy electron. The corresponding pair production rates have been compared with each other in the undercritical ($F'\ll F_{\rm cr}$) and overcritical ($F'\gg F_{\rm cr}$) regimes, considering both linearly and circularly polarized laser fields. To this end, it was in particular necessary to derive an improved analytical formula for the nonlinear Bethe-Heitler rate in the overcritical regime.

We have shown that, for suitably chosen values of the incident-beam Lorentz factors and the bremsstrahlung converter thickness, the nonlinear Bethe-Heitler and Breit-Wheeler rates coincide exactly. In the undercritical regime, the necessary nuclear Lorentz factor is found to be $\approx 1.3$ times larger than the one of the bremsstrahlung-generating electron beam, whereas both Lorentz factors may be chosen equally in the overcritical regime. In both regimes, the adjusted converter thickness exhibits a rather complicated dependence on the various parameters, including the laser field strength. When instead, for reasons of experimental convenience, a fixed bremsstrahlung target thickness is used, the adjustment of the Lorentz factors still allows to obtain good mutual agreement between the nonlinear Bethe-Heitler and Breit-Wheeler rates in a wide range of applied field strengths. We note that the matching was carried out between the Bethe-Heitler rate and the bremsstrahlung-averaged Breit-Wheeler rate per {\it radiating} electron (i.e. the rate that results from the photon spectrum of an incident-beam electron which produces radiation). Alternatively, taking into account that only a fraction $p_\gamma$ of all beam electrons generate bremsstrahlung (e.g., $p_\gamma \approx 1$\% for the parameters of the CALA experiment \cite{CALA}), one could also match the Breit-Wheeler rate per {\it incident} electron with the Bethe-Heitler rate by correspondingly enlarging the bremsstrahlung target thickness from $\ell$ to $\ell/p_\gamma$ (provided the latter remains small against $1$, as required for the applied thin-target approximation).

The total rates of these nonlinear pair production processes -- even though the underlying mechanisms  differ from each other -- can thus be brought into perfect, or at least very close, agreement. Upcoming experiments on the nonlinear Breit-Wheeler process therefore offer the additional benefit that the measured Breit-Wheeler rates can be mapped onto a corresponding Bethe-Heitler scenario with suitably adjusted parameters. In this sense, the bremsstrahlung-driven nonlinear Breit-Wheeler experiments may be considered as to simultaneously 'mimic' the nonlinear Bethe-Heitler process, which hitherto remains unobserved and is experimentally much harder to access. 

\appendix
\section{Derivation of the Bethe-Heitler pair production rate for a linearly polarized laser wave in the overcritical regime}
\label{appendix}

In order to obtain an expression with the desired accuracy for the overcritical  Bethe-Heitler rate in a linearly polarized laser field, we follow the calculational procedure of Ref.~\cite{Milstein}. Similarly as for the Breit-Wheeler process \cite{Landau}, application of the optical theorem allows to obtain the Bethe-Heitler pair production rate in the nuclear rest frame through the imaginary part of the polarization operator $\Pi^{\mu \nu}$ according to \cite{Milstein}

\begin{eqnarray}
\label{W Bethe-Heitler}
\mathcal{R} = 4\pi (Ze)^2\int {d^3q\over (2\pi)^3}{{\rm Im}\,\Pi^{00}\over q^4}.
\end{eqnarray}

For elliptical polarization $\Pi^{00}$ can be taken from Eq.~(6) of Ref.~\cite{Milstein} and depends on the dimensionless intensity parameters $\xi_1$ and $\xi_2$, where the different $\xi_j$ stand for the different polarization directions of the elliptical polarization. In the overcritical regime $\eta_j = \frac{\omega}{m} \xi_j \gg 1$, some of the integrals of the corresponding rate can be evaluated such that only an integral over d$\rho$ is left, which originates from $\Pi^{00}$. A detailed description of the calculation can be found in Ref.~\cite{Milstein} where the resulting rate is given in Eq.~(36). In the case of linear polarization (setting $\xi_1 = \xi$ and $\xi_2 = 0$) it reads

\begin{eqnarray}
	\mathcal{R} &=& {(Z\alpha)^2\omega\over 3\pi^2} \int_0^\infty {d\rho\over\rho^2}
	\Bigg\{ \left[ \ln\left({2 \omega \over m \rho\chi}\right) - C\right] \nonumber\\
	& & \times \Bigg[{\xi^2 \over g_2}(A_1+2\sin^2\rho) - 2\ln{\chi\over 2} \nonumber\\
	& & +{1\over A_0}\left({g_0\over g_2}-1\right)(A_1-3\sin^2\rho)\Bigg] \nonumber\\
	& & +{\xi^2\over g_2}
	\Bigg[A_1\left(\ln{g_0+g_2\over 2g_2} - {4\over 3}\right) \nonumber\\
	& & +2\sin^2\rho\left( \ln{g_0+g_2\over 2g_2} - {19\over 12} \right)\Bigg] \nonumber\\
	& & -\ln{\chi\over 2}\left(\ln{\chi\over 2}-{5\over 3}\right)
	-{1\over 2}L\left({g_2-g_0\over g_2+g_0}\right) \nonumber\\
	& & +{1\over A_0 g_2}\Bigg[ A_1\left(g_0\ln{g_0+g_2\over 2g_2}-{4\over 3}(g_0-g_2)\right) 
	\nonumber\\
	& & -3\sin^2\rho\left(g_0\ln{g_0+g_2\over 2g_2}-{3\over 2}(g_0-g_2)\right)\Bigg]\Bigg\}
\end{eqnarray}

In this expression, one has

\begin{eqnarray}
\label{eq:appendix}
A &=& {1\over 2}\left( 1-{\sin^2\rho\over\rho^2} \right) \approx \frac{\rho^2}{6} , \nonumber\\
A_0 &=& {1\over 2}\left( {\sin^2\rho\over\rho^2}-{\sin 2\rho\over 2\rho} \right) \approx \frac{\rho^2}{6},
\nonumber\\
A_1 &=& A + 2A_0 \approx \frac{\rho^2}{2} , \nonumber\\
g_0 &=& 1 + \xi^2 A \approx 1 + \xi^2 \frac{\rho^2}{6}, \ \ g_1 = \xi^2 A_0 \approx \xi^2 \frac{\rho^2}{6} \nonumber\\
g_2 &=& \sqrt{g_0^2-g_1^2} \approx \sqrt{1+ \frac{\xi^2 \rho^2}{3}}, \nonumber\\
\chi &=& \sqrt{2 \left( g_0 + g_2 \right)} \approx \sqrt{1+ \frac{\xi^2 \rho^2}{3}} + 1.
\end{eqnarray}

Here, the approximations in \eqref{eq:appendix} result from the fact that for $\xi \gg 1$, small values of $\rho$ with $\rho \sim 1/\xi \ll 1$ contribute to the integral. $L$ denotes Spencer's function according to

\begin{eqnarray}
L(x) = \int_0^x{dy\over y}\ln(1+y).
\end{eqnarray}

For small values of $\rho$, we expand the integrand of Spencer's function, such that $L(x) \approx x$.\\
The rate is consequently obtained through
\begin{eqnarray}
	\mathcal{R} &=& {(Z\alpha)^2\omega \xi \over 3 \sqrt{3} \pi^2} \int_0^\infty {d\sigma\over\sigma^2} \nonumber\\
	& & \Bigg\{ \left[ \ln\left({\eta \over 2 \sqrt{3}}\right) - C - \ln\left({\sigma \left(\sqrt{1+\sigma^2} +1 \right) \over 4}\right) \right] \nonumber\\
	& & \times \Bigg[15 - {15 \over \sqrt{1+ \sigma^2}} - 2 \ln\left({\left[ 1 + \sqrt{1+ \sigma^2} \right] \over 2} \right) \Bigg] \nonumber\\
	& & - 23 \left( 1 - {1 \over \sqrt{1+ \sigma^2}} \right) - {\sqrt{1+\sigma^2}-1- \sigma^2 /2 \over 2 \sqrt{1+ \sigma^2} +2 + \sigma^2} \nonumber\\
	& & - \ln \left( {1 + \sqrt{1+ \sigma^2} \over 2} \right) \left[ \ln \left( {1 + \sqrt{1+ \sigma^2} \over 2} \right) - {5 \over 3}\right] \nonumber\\
	& & - {15 \over \sqrt{1+ \sigma^2}} \ln\left({1 + \sigma^2 /2 + \sqrt{1+\sigma^2} \over 2 \sqrt{1+ \sigma^2}} \right) \Bigg\},
\end{eqnarray}
where $\sigma = \frac{\xi \rho}{\sqrt{3}}$. This integral can be calculated analytically and leads to the rate given in Eq.~\eqref{eq: RBHlin_over}, which is expressed in the laboratory frame by virtue of $\mathcal{R}_{\rm BH, lin}^{\rm (oc)} = \frac{1}{\gamma_n} \mathcal{R}$. It is worth noting that the new formula \eqref{eq: RBHlin_over} improves Eq.~(38) from Ref.~\cite{Milstein} for the case of linear laser polarization by including correction terms that go beyond the logarithmic accuracy.



\begin{thebibliography}{33}
	
		\bibitem{Schwinger} J. S. Schwinger, Phys. Rev. {\bf 82}, 664 (1951).
		
		\bibitem{Reiss-1962}
		H.~R.~Reiss, J. Math. Phys. \textbf{3}, 59 (1962).
		
		\bibitem{Nikishov-Ritus-1964}
		A.~I.~Nikishov and V.~I.~Ritus, Zh. Eksp. Teor. Fiz. \textbf{46}, 776 (1963)
		[JETP Lett. \textbf{19}, 529 (1964)].
		
		\bibitem{Ritus-Review}
		V.~I.~Ritus, J. Sov. Laser Res. \textbf{6}, 497 (1985).
		
		\bibitem{Yakovlev} V. Yakovlev, Zh. Eksp. Teor. Fiz. {\bf 49}, 318 (1965)
		[Sov. Phys. JETP {\bf 22}, 223 (1966)].

		\bibitem{Review1} F. Ehlotzky, K. Krajewska, and J. Z. Kami\'nski, 
		Rep. Prog. Phys. {\bf 72}, 046401 (2009).
		
		\bibitem{Review2} R. Ruffini, G. Vereshchagin, and S.-S. Xue, 
		Phys. Rep. {\bf 487}, 1 (2010).
		
		\bibitem{Review3} A. Di Piazza, C. M\"uller, K. Z. Hatsagortsyan, 
		and C. H. Keitel, Rev. Mod. Phys. {\bf 84}, 1177 (2012).
		
		\bibitem{Review4} A. Fedotov, A. Ilderton, F. Karbstein, B. King, 
		D. Seipt, H. Taya, and G. Torgrimsson, Phys. Rep. {\bf 1010}, 1 (2023). 
		
		
		\bibitem{Heinzl2010}
		T.~Heinzl, A.~Ilderton, and M.~Marklund, 
		Phys. Lett. B \textbf{692}, 250 (2010).

		\bibitem{Lebed} A. A. Lebed and S. P. Roshchupkin, Laser Phys. {\bf 21}, 1613 (2011).

		\bibitem{Krajewska-BW}
		K.~Krajewska and J.~Z.~Kami{\'n}ski, Phys. Rev. A \textbf{86}, 052104 (2012).

		\bibitem{Titov2012}
		A.~I.~Titov, H.~Takabe, B.~K\"ampfer, and A.~Hosaka, 
		Phys. Rev. Lett. \textbf{108}, 240406 (2012).

		\bibitem{Krajewska-PRA2013} K. Krajewska, C. M\"uller, and 
		J. Z. Kami\'nski, Phys. Rev. A {\bf 87}, 062107 (2013).

		\bibitem{Meuren-DiPiazza}
		S.~Meuren, K.~Z.~Hatsagortsyan, C.~H.~Keitel, and A.~Di~Piazza, 
		Phys. Rev. D \textbf{91}, 013009 (2015).

		\bibitem{Jansen}
		M. J. A. Jansen and C. M\"uller,
		Phys. Rev. D {\bf 93}, 053011 (2016).
		
		\bibitem{DiPiazzaFocus}
		A.~Di~Piazza, Phys.~Rev.~Lett. \textbf{117}, 213201 (2016).
		
		\bibitem{Heinzl2020} T.~Heinzl, B.~King, and A.~J.~MacLeod, 
		Phys. Rev. A \textbf{102}, 063110 (2020).

		\bibitem{King2021}
		S. Tang and B. King, 
		Phys. Rev. D {\bf 104}, 096019 (2021).		
				
		\bibitem{Krajewska-NJP} K. Krajewska, J. Z. Kaminski, and C. M\"uller,
		New J. Phys. {\bf 23}, 095012 (2021).
		
		
		\bibitem{Narozhny2} N. B. Narozhny and M. S. Fofanov, 
		JETP {\bf 90}, 415 (2000) [Zh. Eksp. Teor. Fiz. {\bf 117}, 476 (2000)].
		
		\bibitem{Loetstedt} E. Lötstedt, U. D. Jentschura, and C. H. Keitel, Phys. Rev. Lett. {\bf 101}, 203001 (2008).
		
		\bibitem{DiPiazza-PRL} A. Di Piazza, E. L\"otstedt, A. I. Milstein, and C. H. Keitel, 
		Phys. Rev. Lett. {\bf 103}, 170403 (2009).

		\bibitem{Krajewska-PRA2012} K. Krajewska and J. Z. Kami\'nski, 
		Phys. Rev. A {\bf 85}, 043404 (2012); Phys. Rev. A {\bf 86}, 021402 (2012).
		
		\bibitem{Augustin} S. Augustin and C. Müller, 
		Phys. Rev. A {\bf 88}, 022109 (2013); Phys. Lett. B {\bf 737}, 114 (2014).
		
		\bibitem{Roshchupkin} S. P. Roshchupkin, N. R. Larin, and V. V. Dubov, Phys. Rev. D {\bf 104}, 116011 (2021).
		
		\bibitem{Mahlin2023}
		N.~Mahlin, S.~Villalba-Ch\'avez, and C.~M\"uller, 
		Phys. Rev. D \textbf{108}, 096023 (2023).	
		
		
		\bibitem{Avetissian} H. K. Avetissian, A. K. Avetissian, G. F. Mkrtchian, 
		and K. V. Sedrakian, Nucl. Instrum. Methods Phys. Res. A {\bf 507}, 582 (2003).
		
		\bibitem{MVG-PRA2003} C. M\"uller, A. B. Voitkiv, and N. Gr\"un, 
		Phys. Rev. A {\bf 67}, 063407 (2003); Phys. Rev. A {\bf 70}, 023412 (2004).		
		
		\bibitem{Sieczka} P. Sieczka, K. Krajewska, J. Z. Kami\'nski, P. Panek, 
		and F. Ehlotzky, Phys. Rev. A {\bf 73}, 053409 (2006).
		
		\bibitem{Krajewska-PRA2006} J. Z. Kami\'nski, K. Krajewska, and F. Ehlotzky, 
		Phys. Rev. A {\bf 74}, 033402 (2006).
		
		\bibitem{Milstein} A. I. Milstein, C. M\"uller, K. Z. Hatsagortsyan, 
		U. D. Jentschura, and C. H. Keitel, 
		Phys. Rev. A {\bf 73}, 062106 (2006).
		
		\bibitem{Kuchiev} M. Y. Kuchiev and D. J. Robinson, Phys. Rev. A {\bf 76}, 012107 (2007).
		
		\bibitem{DiPiazza-PLB} A. Di Piazza and A. I. Milstein, Phys. Lett. B {\bf 717}, 224 (2012).
		
		\bibitem{Titov2020}
		A.~I.~Titov and B.~K\"ampfer, Eur. Phys. J. D \textbf{74}, 218 (2020).
		
		\bibitem{Hatsagortsyan-PRR} F. Wan, Y. Wang, R.-T. Guo, Y.-Y. Chen, 
		R.~Shaisultanov, Z.-F. Xu, K. Z. Hatsagortsyan, C. H. Keitel, and J.-X. Li,
		Phys. Rev. Research 2, 032049(R) (2020).
		
		\bibitem{Riconda} A.~Mercuri-Baron \emph{et al.}, 
		New~J.~Phys. \textbf{23}, 085006 (2021).

		\bibitem{Seipt2020}
		D.~Seipt and B.~King, 
		Phys. Rev. A \textbf{102}, 052805 (2020).
		
		\bibitem{Hatsagortsyan-PRD}
		Y.-Y. Chen, K. Z. Hatsagortsyan, C. H. Keitel, and R.~Shaisultanov,
		Phys. Rev. D 105 116013 (2022).		
		
		\bibitem{Podszus} T.~Podszus, V.~Dinu, and A.~Di~Piazza, 
		Phys. Rev. D \textbf{106}, 056014 (2022).		
		
		\bibitem{Seipt} D. Seipt, M. Samuelsson and T. Blackburn, Plasma Phys. Control. Fusion {\bf 67}, 035002 (2025). 
				

		\bibitem{Serbo} D. Yu. Ivanov, G. L. Kotkin, and V. G. Serbo, 
		Eur. Phys. J. C 40, 27 (2005).
		
		\bibitem{DiPiazza-Spin} A. Di Piazza, A. I. Milstein and C. M\"uller,
		Phys. Rev. A 82, 062110 (2010).
		
		\bibitem{Tim-Oliver} T. O. M\"uller and C. M\"uller, 
		Phys. Lett. B {\bf 696}, 201 (2011);
		Phys. Rev. A {\bf 86}, 022109 (2012).
		
		\bibitem{Selym} S.~Villalba-Ch\'avez and C.~M\"uller, 
		Phys. Lett. B \textbf{718}, 992 (2013).
		
		\bibitem{Jansen-Spin}
		M.~J.~A. Jansen, J.~Z. Kami{\'n}ski, K.~Krajewska, and C.~M\"uller, 
		Phys. Rev. D \textbf{94}, 013010 (2016).		
		
		\bibitem{Tang-Spin} S. Tang,
		Phys. Rev. D 105, 056018 (2022).
		
		\bibitem{Recol1} M. Yu. Kuchiev, Phys. Rev. Lett {\bf 99}, 130404 (2007).
		
		\bibitem{Recol2} S. Meuren, K. Z. Hatsagortsyan, C. H. Keitel, and A. Di Piazza, Phys. Rev. Lett. {\bf 114}, 143201 (2015).
		
		\bibitem{MVG-PRL} C. M\"uller, A. B. Voitkiv, and N. Gr\"un, 
		Phys. Rev. Lett. {\bf 91}, 223601 (2003).
		
		\bibitem{Deneke} C. Deneke and C. M\"uller, Phys. Rev. A {\bf 78}, 033431 (2008).

		\bibitem{diatomic} F. Fillion-Gourdeau, E. Lorin and A. D. Bandrauk, 
		Phys. Rev. Lett. {\bf 110}, 013002 (2013); J. Phys. B {\bf 46}, 175002 (2013).
		
		\bibitem{Grobe-bound} C. K. Li, D. D. Su, Y. J. Li, Q. Su, and R. Grobe, 
		Europhys. Lett. 141, 55001 (2023).

		\bibitem{Grobe-phase}
		C. K. Li, Y. J. Li, Q. Su, and R. Grobe, 
		Phys. Rev. A {\bf 108}, 033112 (2023).
		
		\bibitem{Remme} S. Remme, A. Eckey, S. Villalba-Ch\'avez, A. B. Voitkiv, and C. M\"uller,
		Phys. Rev. A {\bf 113}, 033115 (2026).

		\bibitem{SLAC} D.~L.~Burke \emph{et al.}, 
		Phys. Rev. Lett. \textbf{79}, 1626 (1997).
		
		\bibitem{ELI} I. C. E. Turcu \emph{et al.}, Rom. Rep. Phys. {\bf 68}, 
		S145 (2016); see also \url{https://eli-laser.eu}.
		
		\bibitem{CoReLS} J. W. Yoon \emph{et al.}, Optica \textbf{8}, 630 (2021);
		see also \url{https://corels.ibs.re.kr}.
		
		\bibitem{FACET} S. Meuren, E-320 Collaboration at FACET-II, 
		\url{https://facet.slac.stanford.edu}.
		
		\bibitem{Gemini} C. H. Keitel {\it et al.}, arXiv:2103.06059.

		\bibitem{CALA} F. C. Salgado {\it et al.}, New J. Phys. {\bf 23}, 105002 (2021).
		
		\bibitem{LUXE} H. Abramowicz \emph{et al.}, Eur. Phys. J. Spec. Top. 
		{\bf 230}, 2445 (2021); see also \url{http://www.hibef.eu}.
		
		\bibitem{Reiss1971}
		H.~R.~Reiss, Phys. Rev. Lett. \textbf{26}, 1072 (1971).		

		\bibitem{Blackburn2018}
		T.~G.~Blackburn and M.~Marklund, 
		Plasma Phys. Controlled Fusion \textbf{60}, 054009 (2018).

		\bibitem{Hartin}
		A.~Hartin, A.~Ringwald, and N.~Tapia, Phys.~Rev.~D \textbf{99}, 036008 (2019).

		\bibitem{Eckey2022}
		A.~Eckey, A.~B.~Voitkiv, and C.~M\"uller, Phys.~Rev.~A \textbf{105}, 013105 (2022).

		\bibitem{Golub2022}
		A.~Golub, S.~Villalba-Ch\'avez, C.~M\"uller, Phys.~Rev.~D \textbf{105}, 116016 (2022).
		
		\bibitem{MacLeod} A. J. MacLeod, P. Hadjisolomou, T. M. Jeong, and S. V. Bulanov, Phys. Rev. A {\bf 107}, 012215 (2023).

		\bibitem{King2024}
		B. King and S. Tang, 
		Phys. Rev. A \textbf{109}, 032823 (2024).

		\bibitem{Eckey2024} A.~Eckey, A.~Golub, F.~C. Salgado, S.~Villalba-Ch\'avez, 
		A.~B. Voitkiv, M.~Zepf, and C.~M\"uller, 
		Phys. Rev. A \textbf{110}, 043113 (2024).

		\bibitem{Elsner}
		I. Elsner, A. Golub, S. Villalba-Ch\'avez and C. M\"uller
		Phys. Rev. D {\bf 111}, 096012 (2025).
		
		\bibitem{Jackson} J. D. Jackson, Classical Electrodynamics (John Wiley and Sons, Inc., 1962).
		
		\bibitem{Landau} V. B. Berestetskii, E. M. Lifshitz, and L. P. Pitaevskii, 
		Relativistic Quantum Theory (Pergamon Press, London, 1971).
		
		\bibitem{bremsstrahlung1} P. A. Zyla {\it et al.} (Particle Data Group), 
		Rev. Mod. Phys. \textbf{2020}, 083C01 (2020).
		
		\bibitem{bremsstrahlung2} Y.-S. Tsai, 
		Rev. Mod. Phys. \textbf{46}, 815 (1974).
		
		\bibitem{Narozhny} N. B. Narozhny, Phys. Rev. D {\bf 21}, 1176 (1980).
		
		\bibitem{Seipt2025} N. Larin and D. Seipt, Phys. Rev. A {\bf 112}, 032819 (2025).
		
		\bibitem{STAR} In connection, we note that the linear Breit-Wheeler process (Eq.~\eqref{BW} with $n=1$) has been observed in ultraperipheral heavy-ion collisions, where it is induced by quasi-real photons from the relativistically moving ions; see J. Adam {\it et al.} (Star Collaboration), Phys. Rev. Lett. 127, 052302 (2021).

	\end{thebibliography}
\end{document}